\documentclass{cargese}\usepackage{graphicx}

\let\footnote\savefootnote
\let\footnotetext\savefootnotetext 
\let\footnoterule\savefootnoterule 
\normallatexbib

\newcommand{\beqn}{\begin{eqnarray}}
\newcommand{\eeqn}{\end{eqnarray}}
\newcommand{\be}{\begin{equation}}
\newcommand{\ee}{\end{equation}}
\newcommand{\non}{\nonumber \\}
\newcommand{\CYT}{$Y_{\mbox{\scriptsize 3}}$}
\newcommand{\CYF}{$Y_{\mbox{\scriptsize 4}}$}
\newcommand{\vol}{{\cal V}}

\newcommand{\bj}{\bar{\jmath}}
\newcommand{\pu}{\partial_{\mu}}
\newcommand{\po}{\partial^{\mu}}
\newcommand{\fn}{\footnotesize}

\renewcommand{\theequation}{\arabic{equation}}
\begin{document}

\articletitle[3D Heterotic/M-Theory Duality]{Aspects of
  Heterotic/M-Theory\\ Duality in D=3}

\chaptitlerunninghead{3D Heterotic/M-Theory Duality}

\author{Michael Haack\footnote{%
Presented by M.\ Haack.} and Jan Louis}

\affil{Fachbereich Physik, Martin-Luther-Universi\"at Halle-Wittenberg\\
       Friedemann-Bach-Platz 6, D-06108 Halle, Germany}         

\begin{abstract}
We study the duality between M-theory compactified on 
Calabi-Yau fourfolds and the
heterotic string compactified on Calabi-Yau threefolds
times a circle.
 Our analysis is based on a comparison of the low energy effective actions in three dimensions.
\end{abstract}

Non-perturbative N=1 vacua of the 
heterotic string in D=4
are of particular interest due to their
phenomenological prospects \cite{KL}.
A certain class of such vacua is best described
by F-theory  compactified on elliptic 
Calabi-Yau fourfolds \CYF\, \cite{V,L}.
A closely related but technically simpler
class of vacua is obtained by further
compactifying the heterotic string on a circle $S^1$
down to D=3; some of their
non-perturbative properties 
are captured by M-theory compactified on \CYF\,
\cite{BB}.
In this short note we display the corresponding
effective Lagrangians and their relation
following a similar analysis carried out
for the duality of type IIA on \CYT\ and the 
heterotic string on K3$\times T^2$ in 
ref.\ \cite{AL}.
A more detailed presentation of our results will be given in a forthcoming paper \cite{HL}.

Our starting point is a generic
effective Lagrangian of N=1, D=4 heterotic string 
vacua\footnote{They  can be
  constructed as compactifications on \CYT\
 or more generally from appropriate (0,2) SCFTs.}
(using the conventions of ref.\ \cite{FL})
\beqn
{\cal{L}}^{(4)} & = & \frac{R}{2} - G_{\bar{i}j}^{(4)} 
\partial_m
\bar{\Phi}^{\bar i} \partial^m \Phi^{j} 
- \frac{1}{4} \mbox{Re} f_{a}(\Phi) F^{a}_{mn} F^{amn} \non
& & \mbox{} + \frac{1}{4} \mbox{Im} f_{a}(\Phi) 
F^{a}_{mn} \tilde F^{amn}+ \ldots, 
\label{eq1}
\eeqn
where  $m,n=0,\ldots,3$.
$\Phi^i$ are moduli fields including the (1,1) and (1,2)
moduli of \CYT\ as well as moduli arising from the choice 
of the gauge bundle.\footnote{We ignore all charged matter
multiplets and solely focus on the chiral moduli 
multiplets and the vector multiplets. We also
neglect
the possibility of anomalous $U(1)$ gauge factors with appropriate
4-dimensional Green-Schwarz terms.}
$G_{\bar{i}j}^{(4)}$ is the K\"ahler metric which is determined by a K\"ahler
potential $K^{(4)}(\Phi,\bar{\Phi})$ and the 
$F^a_{mn}$ denote the field strengths of the (non-Abelian)
gauge bosons $A^a_{m}$. 
The $S^1$-reduction to $D=3$ uses the Ansatz \cite{FS}:
\be
g^{(4)}_{mn} = \left( \begin{array}{cc}
                 g^{(3)}_{\mu \nu} + e^{2 \sigma} B_{\mu} B_{\nu} & e^{2 \sigma} B_{\mu}\\
                 e^{2 \sigma} B_{\nu} & e^{2 \sigma}
                 \end{array}
          \right),
\hspace{.2cm} A^{a}_{m} = \left(A^{a}_{\mu} + B_{\mu} \zeta^{a}, \zeta^{a} \right), \label{eq2}
\ee
where $\mu,\nu = 0,1,2$.
In D=3 the vector multiplet contains an adjoint scalar $\zeta^{a}$ so that the
gauge group $G$ is generically broken to its 
Abelian subgroup $[U(1)]^{\mbox{r}(G)}$. 
(By slight abuse of notation we choose 
to label in the following the 
different $U(1)$ gauge multiplets by the index $a$  i.e.\, 
$a=1,\ldots,\mbox{r}(G)$). 
In D=3 an Abelian vector is dual to a scalar 
and we denote by $C^{a}$
the scalars dual to $A^{a}_\mu$ and by 
$\tilde{\Phi}$ the dual scalar of $B_{\mu}$. 
In the dual picture all supermultiplets are chiral
and their scalar fields 
parametrize a K\"ahler manifold \cite{DNT}. 
The K\"ahler structure becomes manifest 
in the coordinates
\beqn
D_{a} & = & -f_{a}(\Phi) \zeta_{a} + i C_{a}\ ,  \label{eq3} \\
T & = & e^{2 \sigma} + i \tilde{\Phi} 
+ \frac{1}{2} (\mbox{Re}f_{a}(\Phi))^{-1}\,
D^{a}(D^{a} + \bar{D}^{a})  \  .  \nonumber 
\eeqn
Inserting (\ref{eq2}) into (\ref{eq1}) using (\ref{eq3}),
performing the duality transformation and a Weyl rescaling
results in the 3-dimensional K\"ahler potential
\be
K^{(3)}_{\mbox{\footnotesize het}} = 
K^{(4)}(\Phi,\bar{\Phi}) 
- \ln \left[ T + \bar{T} - \frac{1}{2} (D^{a} + \bar{D}^{a})^{2} (\mbox{Re} f_{a}(\Phi))^{-1} \right].  \label{eq4b}
\ee
This can be further simplified by 
using the known tree level form of $K^{(4)}$
and $f_a$. One has $f_{a} = S$ and 
$K^{(4)} = \tilde{K}^{(4)} - \ln(S+\bar{S})$
where $\tilde{K}^{(4)}$ depends on all
moduli but the 4-dimensional dilaton $S$ \cite{FL}.
For this case (\ref{eq4b}) reads
\be
K^{(3)}_{\mbox{\footnotesize het}} = \tilde{K}^{(4)} 
- \ln \left[(T+\bar{T})(S+\bar{S}) - (D^{a} + \bar{D}^{a})^{2} \right].  \label{eq5}
\ee

On the M-theory side we start with the 11-dimensional supergravity Lagrangian \cite{CJS}:
\be
{\cal L}^{(11)} = \frac{1}{2} R - \frac{1}{4}
  |F_{4}|^{2} - \frac{1}{12} A_{3} \wedge F_{4} \wedge F_{4} \ , \label{eq6}
\ee
where $A_3$ is a 3-form and $F_4$ its field strength.
A Kaluza-Klein reduction on \CYF \, in the spirit of
\cite{BCF} leads to massless bosons in D=3
corresponding to  the Hodge-numbers
$h_{1,1}, h_{1,2}$ and $h_{1,3}$ of 
\CYF.\footnote{We choose to focus on such
\CYF\ that lead to 
duals of the perturbative heterotic string
without anomalous U(1) factors. They obey 
$\chi= 0$ and do not have non-trivial 4-form flux 
or space-time filling membranes in the vacuum.} 
The corresponding harmonic forms are denoted by 
$V^{A}\in H^{1,1}, \Psi^{I} \in H^{2,1}$ and
$\Phi^{\alpha} \in H^{3,1}$.
The K\"ahler  and complex structure deformations of 
the Calabi-Yau metric
$g_{i\bar{j}}$ are \cite{BCF}
\be\label{eq8}
i \delta g_{i \bar{j}} 
= \sum_{A=1}^{h_{1,1}} \tilde{M}^{A} V^{A}_{i \bar{j}}\
,
\qquad \delta g_{\bar i \bar j} = \sum_{\alpha=1}^{h_{3,1}} 
{Z}^{ \alpha}
{b}^{ \alpha}_{\bar i \bar j}\ , 
\ee
where ${b}^{\alpha}_{\bar i \bar j}$ is related to
$\Phi^{\alpha}$ by an appropriate contraction with
the anti-holomorphic 4-form $\bar\Omega$ on \CYF.
The 3-form $A_3$ is expanded accordingly
\be
A_{3} = A_{1}^{A} \wedge V^{A} + n^{I} \Psi^{I} + \bar{n}^{\bar{J}} \bar{\Psi}^{\bar{J}},  \label{eq9}
\ee
where $A_{1}^{A}$ are 3-dimensional 1-forms which
can again be dualized to scalars $N^{A}$. 
Inserting (\ref{eq8}) and (\ref{eq9}) into
(\ref{eq6}) results in the effective Lagrangian 
\beqn
{\cal L}^{(3)} & = & \frac{R}{2}  - G_{\bar{\alpha} \beta} \partial_{\mu}
\bar{Z}^{\bar \alpha} \partial^{\mu} Z^{\beta} - 
G_{I\bar{J}} D_{\mu} n^{I} D^{\mu} \bar{n}^{\bar{J}} -\frac{1}{2} G_{AB} \partial_{\mu} M^{A} \partial^{\mu} M^{B}  \non
& & -\frac{1}{32\vol} 
\Big(4 \pu N^{A} + \kappa^{A K \bar{L}} 
(n^K D_{\mu} \bar{n}^{\bar{L}} - D_{\mu}  n^{K} 
\bar{n}^{\bar{L}}) \Big) G^{-1}_{AB} \non
& & \hspace{1.1cm} \Big(4\po N^{B} + \kappa^{B I \bar{J}} 
( n^{I} D^\mu \bar{n}^{\bar{J}} - D^\mu n^{I} \bar{n}^{\bar{J}}) \Big), \label{eq10}  
\eeqn
where we rescaled 
$\tilde M^{A} = \vol^{-1/6}{M}^{A}$ with 
$\vol \equiv \frac{1}{4!} d_{ABCD} \, M^{A}M^{B}M^{C}M^{D}$ 
and $d_{ABCD}$ being the intersection numbers of \CYF.
Furthermore we used the following definitions
in close analogy with Calabi-Yau 
threefolds \cite{BCF,S,CO}
\beqn
\kappa^{A I \bar{J}} & \equiv & \int_{\mbox{\footnotesize \CYF}} V^{A} \wedge
\Psi^{I} \wedge \bar{\Psi}^{\bar{J}} \ ,  \non
G_{I \bar{J}} & \equiv &  \frac{1}{2\vol^{1/3}}
\int_{\mbox{\footnotesize \CYF}} \Psi^{I} \wedge \star {\Psi}^{{J}} 
= - \frac{1}{2} i \kappa^{A I \bar{J}} M^{A} {\vol}^{-1/2} \ , \label{eq12} \\
G_{AB} & \equiv & \frac{1}{2\vol} 
\int_{\mbox{\footnotesize \CYF}} V^{A} \wedge \star V^{B}\ ,\qquad 
G_{\bar{\alpha} \beta} \equiv \frac{1}{4\vol^{1/3}} 
\int_{\mbox{\footnotesize \CYF}} d^{8} \xi
  \sqrt{g} \bar{b}_{\bar \alpha}^{\bj \bar{m}} b_{\beta \bj \bar{m}}.  
\nonumber
\eeqn
$G_{\bar{\alpha} \beta}$ is 
the K\"ahler metric of the 
K\"ahler potential $K= -\ln \left( \int_{\mbox{\CYF}} \Omega \wedge 
\bar{\Omega} \right)$ \cite{AS}.
We also abbreviated
\be
D_\mu n^I = \pu n^I + n^K \Theta_{\alpha K I} 
\, \pu Z^\alpha + 
\bar{n}^{\bar{L}} 
\bar{\tilde{\Theta}}_{\bar{\beta} \bar{L} I} 
\, \pu \bar{Z}^{\bar{\beta}}, \qquad 
D_\mu \bar{n}^I = \overline{D_\mu n^I}, 
\ee
where the $\Theta$ and $\tilde{\Theta}$ are (unknown) functions 
of $Z^\alpha, \bar{Z}^{\bar\alpha}$
defined as the coefficient functions of 
\be
\pu \Psi^I = \left( \Theta_{\beta I K}(Z^\alpha, \bar{Z}^{\bar\alpha}) 
\Psi^K + 
\tilde{\Theta}_{\beta I \bar{L}}(Z^\alpha, \bar{Z}^{\bar\alpha}) 
\bar{\Psi}^{\bar{L}} \right) \pu Z^\beta.
\ee 
So far we have not been able to find a 
K\"ahler potential for all the scalars in (\ref{eq10}). 
However if we freeze the complex structure moduli 
and define 
\be
\zeta^{A} = \frac{1}{\sqrt{8}} 
\Big( i N^{A} + G_{AB} \vol^{1/2} M^{B} -\frac{i}{4} \kappa^{A I \bar{J}}
n^{I} \bar{n}^{\bar{J}} \Big)
\ .  \label{eq16}
\ee
the K\"ahler potential for the metric in terms of 
$(\zeta^A,n^I)$ is found to be 
\be
\tilde{K}^{(3)}_{\mbox{\fn M}} = 
- \ln\Big[(\zeta^{A} + \bar{\zeta}^{A} +
\frac{i \kappa^{A  I \bar{J}}}{2 \sqrt{8}}  n^{I} \bar{n}^{\bar{J}}) G^{-1}_{AB} ( \zeta^{B} +
\bar{\zeta}^{B} +\frac{i \kappa^{B K \bar{L}}}{2 \sqrt{8}}  n^{K}
\bar{n}^{\bar{L}})\Big] \ \label{eq17}. 
\ee

Equating the full K\"ahler potentials 
$K^{(3)}_{\mbox{\fn M}}= K^{(3)}_{\mbox{\footnotesize het}}$ 
results in a condition
for the intersection numbers $d_{ABCD}$.
In the following we show the equivalence
of the two K\"ahler potentials in a particular 
limit and
postpone  the general analysis to ref.\ \cite{HL}. 
Fibering the 7-dimensional duality (M/K3 $\simeq$ Het/T$^{3}$) \cite{W}
suggests to choose a K3 fibred fourfold 
(where we assume that there are no bad
fibres introducing additional (1,1)-forms) and an elliptically fibred
threefold \cite{L}. 
As the base $B_2$ we take in both theories
a Hirzebruch surface (whose
two K\"ahler moduli we denote by $U$ and $V$). 
Furthermore we freeze the values of the scalars 
$n^{I}, Z^\alpha$ to zero and consider on both sides the large
base limit. In this limit 
one can choose the divisors of  \CYF\, 
in such a way, that the leading terms of
$K^{(3)}_{\mbox{\fn M}}$ 
correspond to the intersection numbers  
$d_{UV\hat{A}\hat{B}} = \eta_{\hat{A}\hat{B}} = (+,-, \dots, -)$, 
where $\hat{A},\hat{B}$ denote the K\"ahler moduli of the generic
K3 fibre and $\eta_{\hat{A}\hat{B}}$ is the intersection 
form of the corresponding divisors in K3. 
Using these intersection numbers in (\ref{eq17}) results in
\be
K^{(3)}_{\mbox{\fn M}}
 = - \ln\Big[(\zeta^{U} + \bar{\zeta}^{U}) (\zeta^{V} + \bar{\zeta}^{V}) (\zeta^{\hat{A}}
 + \bar{\zeta}^{\hat{A}}) \eta_{\hat{A}\hat{B}} (\zeta^{\hat{B}} + \bar{\zeta}^{\hat{B}})\Big] \ .  \label{eq18}
\ee
On the heterotic side we 
take only $S,T,D^a$ and the
(1,1) moduli $U,V$ of the base $B_2$ into account and
freeze all other (1,1), all (1,2) and all gauge bundle
moduli.
In the limit of a large base $\tilde{K}^{(4)}$
simplifies and one has
$\tilde{K}^{(4)} = -\ln (U+\bar U)(V+\bar V)$.
Thus, the K\"ahler potentials agree if one
identifies
$(S,T,D^{a}) \leftrightarrow (\zeta^{\hat{A}})$ and 
$(U,V) \leftrightarrow (\zeta^{U}, \zeta^{V})$. 
This result is in accord with the formulas 
relating the spectra of dual pairs given in \cite{L,M}.

\begin{acknowledgments}
We would like to thank T. Bauer, B. Hunt, W. Lerche, D. L\"ust, P. Mayr, A. Sen, S. Stieberger and A. Strominger for useful discussions. M.H. would also like to thank the organizers of Carg\`ese '99 ASI for a very nice school and for financial support. 
\end{acknowledgments}


\end{document}